\documentclass[%
preprint,aps,prb,nofootinbib,superscriptaddress
]{revtex4-2}

\usepackage{graphicx}
\usepackage{dcolumn}
\usepackage{bm}
\usepackage{lipsum}
\usepackage{bm} 
\usepackage[thinc]{esdiff} 
\usepackage{amsmath} 

\usepackage{textcomp}
\usepackage{appendix}

\begin{document}

\title{Magnetic properties and growth kinetics of Co/Gd bilayers with perpendicular magnetic anisotropy}
\author{Thomas~J.~Kools}
\email{t.j.kools@tue.nl}
\affiliation{Department of Applied Physics, Eindhoven University of Technology, P.O. Box 513, 5600 MB Eindhoven, the Netherlands}
\author{Julian~Hintermayr}
\affiliation{Department of Applied Physics, Eindhoven University of Technology, P.O. Box 513, 5600 MB Eindhoven, the Netherlands}
\author{Youri~L.\,W.~van~Hees}
\affiliation{Department of Applied Physics, Eindhoven University of Technology, P.O. Box 513, 5600 MB Eindhoven, the Netherlands}
\author{Kenneth~Poissonnier}
\affiliation{Department of Applied Physics, Eindhoven University of Technology, P.O. Box 513, 5600 MB Eindhoven, the Netherlands}
\author{Mark~C.\,H.~de~Jong}
\affiliation{Department of Applied Physics, Eindhoven University of Technology, P.O. Box 513, 5600 MB Eindhoven, the Netherlands}
\author{Jens~T.\,J.\,M.~Janssen}
\affiliation{Department of Applied Physics, Eindhoven University of Technology, P.O. Box 513, 5600 MB Eindhoven, the Netherlands}
\author{Bert~Koopmans}
\affiliation{Department of Applied Physics, Eindhoven University of Technology, P.O. Box 513, 5600 MB Eindhoven, the Netherlands}
\author{Reinoud~Lavrijsen}
\affiliation{Department of Applied Physics, Eindhoven University of Technology, P.O. Box 513, 5600 MB Eindhoven, the Netherlands}

\date{\today}


  \newpage

	\begin{abstract}
    Ultrathin 3d-4f synthetic ferrimagnets with perpendicular magnetic anisotropy (PMA) exhibit a range of intriguing magnetic phenomena, including all-optical switching of magnetization (AOS), fast current-induced domain wall motion (CIDWM), and the potential to act as orbital-to-spin angular momentum converters. For spintronic applications involving these materials, the Curie temperature is a crucial factor in determining not only the threshold energy for AOS, but also the material's resistance to temperature rise during CIDWM. However, the relationship between the Curie temperature, the thicknesses of the individual layers, and the specifics of the growth process remains an open question. In this work, we thoroughly investigate the Curie temperature of one of the archetype synthetic ferrimagnets with PMA, the Pt/Co/Gd trilayer, grown by DC magnetron sputtering and characterized with MOKE and SQUID. We provide an interpretation of the experiments we designed to address these outstanding questions through modeling of the deposition process and the induced magnetization at the Co/Gd interface. Our findings demonstrate that the Curie temperature and, by extension, the conditions for PMA and magnetic compensation, of these ultrathin 3d-4f synthetic ferrimagnets are not only impacted by the interface quality, which can be influenced by the sputtering process, but also to a significant extent by finite-size effects in the 4f-material. This work offers new methods and understanding to predict and manipulate the critical temperature and magnetostatic properties of 3d-4f synthetic ferrimagnets for spintronic applications and magneto-photonic integration.
	\end{abstract}

\maketitle
 
	\section{Introduction}
		Ferrimagnetic spintronics is an active research field that aims to combine the fast magnetization dynamics and low intrinsic magnetization of antiferromagnets with the ease of manipulation and characterization of ferromagnetic materials \cite{Kim2022,Sala2022,Zhang2023}.  Among the large family of ferrimagnetic materials, certain alloys and multilayers composed of 3d-ferromagnets (such as Co, Fe, Ni) and 4f-ferromagnets (such as Gd, Tb, Ho) are of particular interest due to their ability to exhibit single-pulse all-optical switching of magnetization (AOS) \cite{Stanciu2007,Radu2011,Ostler2012,Mangin2014,Lalieu2017,vanHees2020,Li2022He,Hintermayr2023AOS,Verges2023}. Additionally, these systems have garnered significant attention from the scientific community for their effective spin-orbit torque (SOT)-driven manipulation of magnetic order~\cite{Mishra2017,Ueda2017,Je2018,Sala2022Switch}, the presence and efficient motion of skyrmions~\cite{Wang2022,Quessab2022,Brandao2019, Caretta2018,Mallick2024,Hassan2024}, the hosting of optically excitable terahertz-scale exchange resonances at room temperature~\cite{Hintermayr2023EXM, Hintermayr2024}, and exchange torque-driven current-induced domain wall motion with velocities exceeding 1000 m/s~\cite{Caretta2018,Cai2020,LiKools2023}. The combination of these phenomena in a single material system makes them a promising candidate to bridge the gap between photonics and spintronics~\cite{Lalieu2019,Wang2022MTJ,LiKools2023,Pezeshki2023}.

        Within this class of materials, a distinction is often made between alloys, which have an approximately uniform distribution of TM and RE elements, and synthetic ferrimagnets, which consist of discrete layers of the RE and TM. The latter category has several distinct advantages. Contrary to alloys with PMA, a much wider composition range between the 3d and 4f-metal exhibits single-pulse AOS \cite{Beens2019Alloy,Beens2019Intermixing}. Combined with increased access to interfacial engineering, this leads to greater flexibility and tunability of their magnetic properties. Moreover, the Pt/Co/Gd trilayer displays strong interfacial effects, such as perpendicular magnetic anisotropy (PMA), the spin-Hall effect \cite{Haazen2013}, and the interfacial Dzyaloshinskii–Moriya interaction \cite{Cao2020,Hartmann2019,Pham2016}, which are essential ingredients for applications based on efficient domain wall motion or SOT-driven manipulation \cite{Blasing2020,Sala2022}. Furthermore, 3d/4f synthetic ferrimagnets have also been found to be potentially efficient orbital-to-spin angular momentum converters \cite{Sala2022Hall}.

        Despite significant progress in demonstrating the applicability of this material system for future spintronic devices, several important questions remain unaddressed regarding the basic magnetic properties of this heterogeneous 3d-4f system, particularly with respect to Curie temperature ($T_\text{C}$). $T_\text{C}$ has been identified as a key quantity for the energy efficiency of AOS \cite{Li2022He,Igarashi2020}, and for describing current-induced domain wall motion (CIDWM) under the influence of Joule heating \cite{LiKools2023}. These are two cornerstones for potential future magneto-photonic memory applications \cite{Pezeshki2023,Pezeshki2023IEEE,Pezeshki2024}. Furthermore, by unraveling the relationship between $T_\text{C}$ and the Gd thickness, we deepen our understanding of magnetostatic properties such as magnetic compensation and effective anisotropy, important properties for applications.

\begin{figure}
			\centering
			\includegraphics{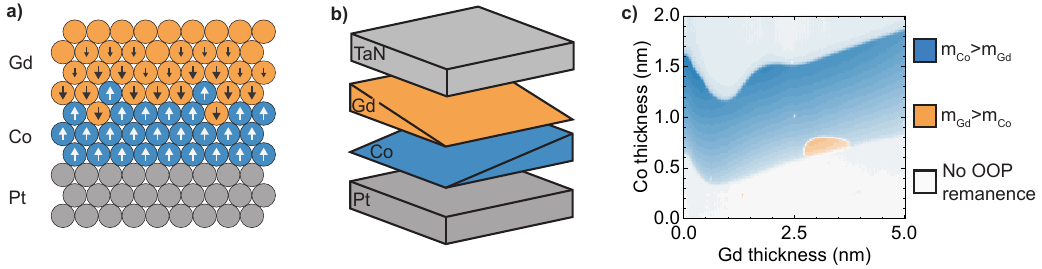}
			\caption{a): Schematic of the (proximity-induced) magnetization of the Pt/Co/Gd trilayer stack. b): Schematic of the double wedge samples under discussion in this work. c): PMOKE remanence scan on a Pt/\-Co(0-2)/\-Gd(0-5) sample.}
			\label{fig:fig1}
		\end{figure}

        Earlier research has indicated that one of the key characteristics of these synthetic ferrimagnets is their interfacially induced magnetization, as schematically illustrated in Fig. \ref{fig:fig1}a. Gd with a thickness of the order of a few nanometers, as considered in this work, has a $T_\text{C}$ well below room temperature. This leads to its magnetization being primarily localized at the Co/Gd interface due to the Co-Gd exchange \cite{Basha2018,brandão2024}. This suggests an important dependence of the net magnetization on the interface, with a more (less) intermixed interface between Co and Gd leading to a larger (smaller) amount of magnetization being induced in the Gd. Fingerprints of this effect have been observed in earlier work by Wang \textit{et al.}. \cite{Wang2020}, who noted a marked decrease in net magnetization after annealing. This decrease was attributed to the phenomena of intermixing at the Co / Gd interface, which highlights the relevance of the interface, as also noted in earlier experimental studies of similar materials \cite{Alonso2002,Andres2000,Colino1999,BASHA2021, Singh2022, GONZALEZ2004, Gonzalez2002}. 
        Moreover, in the ultrathin film regime, where $T_\text{C}$ of the films starts to be influenced by finite-size effects, and for systems consisting of two layered materials with different ordering temperatures, the nature and temperature of the phase transition are strongly influenced by the relative thickness of the materials and their coupling\cite{Srivastava1998,Wang1992,Willis2007,Bovensiepen1998}. However, a comprehensive study and overview of the influence of these and other effects on $T_\text{C}$, the presence of PMA, and the magnetic compensation of 3d-4f synthetic ferrimagnets is still absent.
		
         In this work, we provide a detailed overview of the physical effects that can be considered to predict and model the experimental behavior of one of the simplest model 3d-4f systems with PMA, the Pt/Co/Gd trilayer structure. This study is divided into two parts. First, we present an extensive experimental investigation of the Co/Gd system using the polar magneto-optical Kerr effect (PMOKE), longitudinal MOKE (LMOKE), and vibrating sample magnetometry with a superconducting quantum interference device (VSM-SQUID). Subsequently, we introduce two different models to interpret the observed response: one based on deposition kinetics to simulate sputter-induced intermixing and an analytical model of the induced magnetic moment at the Co/Gd interface. Ultimately, we demonstrate that $T_\text{C}$, and by extension the conditions for PMA and magnetic compensation, are strongly influenced by the thickness of Gd, inducing changes of up to 150 K in $T_\text{C}$. This occurs even at Gd thicknesses where the interface between the Co and Gd is no longer affected while maintaining the PMA. These findings open up new ways for manipulating and engineering the magnetostatic properties of this promising material system for magneto-photonic and spintronic applications.
         
    \section{Experiments}
    \subsection{Methods}
		The samples characterized in this work were grown by DC magnetron sputtering on Si substrates with 100 nm of thermal oxide. The general stack structure is Ta(4)/\-Pt(4)/\-Co($t_\text{Co}$)/\-Gd($t_\text{Gd}$)/\-TaN(4)/\-Pt(2), where the numbers in parentheses indicate layer thicknesses in nanometers. In addition to full sheet samples with uniform layer thickness, we also grew samples with thickness variations across 20 mm (lateral dimensions) using a wedge shutter that gradually closed during the deposition process. The sputter deposition system operated at a typical base pressure of $5\cdot10^{-9}$ mBar, with other details of the deposition system described elsewhere \cite{Kools2023}.  MOKE characterization was performed at room temperature ($T_\text{room}$) using a laser diode with a wavelength of 658 nm. We assume sensitivity only to the magnetization components of Co, as Gd does not significantly contribute to the MOKE signal at this wavelength \cite{Erskine1973}. In-plane (IP) and out-of-plane (OOP) VSM-SQUID characterization was performed using a QuantumDesign MPMS 3 system. These SQUID measurements were performed on full sheet samples, which were diced into approximately $4.5 \times 4.5$ mm$^2$ pieces.

    \subsection{MOKE characterization}
        We begin our discussion of this material system by examining a sample schematically shown in Fig. \ref{fig:fig1}b, which features two orthogonally wedged layers.
       These samples are created by rotating the sample 90$^\circ$ between the deposition of Co and Gd. Subsequently, we saturate the sample with an OOP magnetic field of 1 T and scan the sample surface to determine the remanence from the PMOKE signal at each point, in the absence of an applied magnetic field. This method allows us to scan a wide range of nominal layer thickness combinations of Co and Gd in a single sample. A characteristic result of such a 2D-wedge scan is shown in Fig. \ref{fig:fig1}c for a Ta(4)/\-Pt(4)/\-Co(0-2)/\-Gd(0-5)/\-TaN(4)/\-Pt(2) sample. 

       This static characterization reveals a varied response depending on the thicknesses of the two magnetic layers. We first focus on the different colored regions in the phase diagram in Fig. \ref{fig:fig1}c, as they define the regions and boundaries of interest. At the origin in Fig. \ref{fig:fig1}c, we find no magnetic material and hence no remanence in the PMOKE signal, indicated by the white color. As we increase the Co-thickness, $t_\text{Co}$, vertically in the phase diagram, the system eventually transitions to a magnetized state with an OOP orientation (dark blue) due to the perpendicular magnetic anisotropy (PMA) from the Pt/Co interface. For even larger $t_\text{Co}$, the dipole energy starts to dominate, and the magnetization transitions to an in-plane orientation, which leads to no contrast in the PMOKE remanence signal (white again). As a function of the Gd thickness, the boundaries between these different regions shift to different Co thicknesses. Notably, within the OOP region in the phase diagram, there is a small orange pocket where the magnetization of the Co points in the direction opposite to the saturating field, indicating that the Gd moment is larger in this region. This is an interesting observation, as previous work suggested that the small interfacial Gd magnetization would always be dominated by Co magnetization \cite{Lalieu2017,Pham2016,Bläsing2018}. We find that the capping layer, TaN, plays an important role in facilitating a larger total moment in Gd, consistent with our previous results \cite{Kools2023,Kools2022}. It is again important to note that the interfacial nature of Gd magnetization follows from the Co-Gd exchange, given that the Curie temperature of Gd thin films is significantly lower than its bulk value (295 K) \cite{Farle1993,Jiang1995,Zhang2001}.

    \begin{figure}
			\centering
			\includegraphics{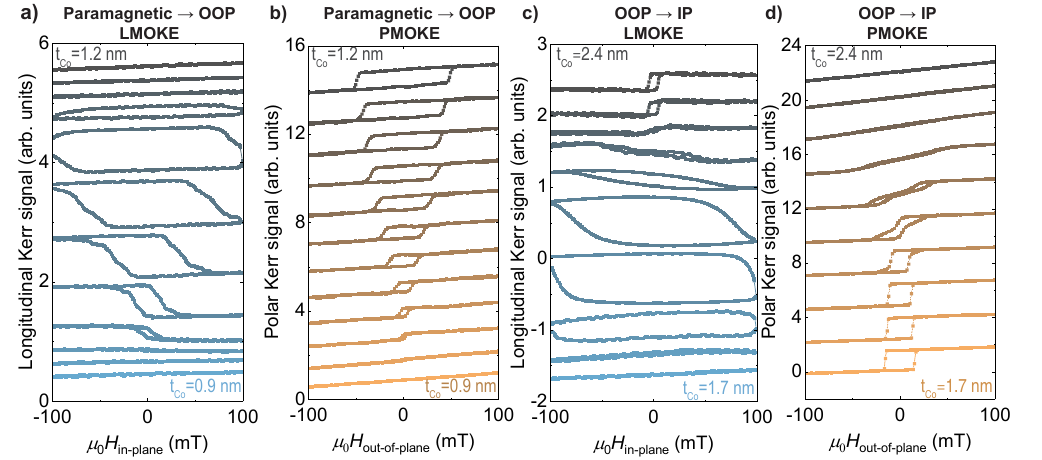}
			\caption{a,b): MOKE response around $t_\text{OOP}$ as a function of Co-thickness in a Co(x)/Gd(10) sample in the longitudinal (a) and polar (b) geometry. c,d): MOKE response around $t_\text{IP}$ as a function of Co-thickness in a Co(x)/Gd(10) sample in the longitudinal (c) and polar (d) geometry.}
			\label{fig:fig2}
		\end{figure}
       
       The appearance of these different regimes (paramagnetic, OOP, IP) as a function of Co thickness reflects the typical competition between the interfacial OOP anisotropy from the Pt/Co interface and the dipole energy. However, the main interest here is what happens to the boundaries separating these two regimes with increasing $t_\text{Gd}$. For this purpose, it is convenient to define the Co thicknesses at which the transitions from paramagnetic to OOP and from OOP to IP magnetization take place, $t_\text{OOP}$ and $t_\text{IP}$, respectively.

       We broadly distinguish two trends in $t_\text{OOP}$ and $t_\text{IP}$: a decrease at $t_\text{Gd}\lesssim0.8$ nm, and an increase across the remaining range on this 2D-wedge. We will restrict ourselves to a qualitative discussion of the possible origins of these trends here and dive into the quantitative expected contributions of these later on when we discuss our modeling efforts. Starting with the initial decrease of $t_\text{OOP}$, we speculate that the main contributing factor is the replacement of the neighboring layers of Co from TaN to Gd. Ta is known to induce a magnetic dead layer in typical 3d ferromagnets \cite{Kowalewski2000,Sato2016}, and although this effect has not been consistently studied for TaN, the additional exchange energy generated by replacing the Co/Ta interface by a Co/Gd interface can be expected to increase the overall exchange energy for the same Co thickness, leading to a higher ordering temperature. However, the decrease in $t_\text{IP}$ is surprising, given the expected decrease in net magnetization with the introduction of Gd and the typical scaling of the dipole energy with the square of the net magnetization. We speculate that the decrease in effective anisotropy is related to the (partial) amorphization of Co upon Gd deposition \cite{Alonso2002,Quiros2008,Quiros2012}, which decreases the interfacial anisotropy at the (111)-textured Pt/Co interface. Additionally, there is a local maximum of $t_\text{IP}$ at around 2.2 nm. Diving into this is beyond the scope of this work, but we note that the (damped) oscillatory behavior may be a fingerprint of the RKKY-like exchange in pure Gd \cite{Scheie2022,brandão2024} and may deserve further study.

        Above $t_\text{Gd} = 0.8$ nm, we see a more pronounced effect on both $t_\text{OOP}$ and $t_\text{IP}$, as both thicknesses increase for the entire thickness range up to 5 nm, and continue to do so up to approximately 11 nm (see Appendix \ref{app:Longrangewedge}). Although the signal magnitude decreases due to stronger attenuation of the MOKE signal from Co by the Gd on top, the trends in $t_\text{IP}$ and $t_\text{OOP}$ can still be observed. These trends may be considered a surprising result given that the magnetization in the Gd is mostly localized at the Co/Gd interface, and it is not expected that the Gd deposited at these larger thicknesses exhibits a significant magnetic order \cite{brandão2024}. 

\begin{figure}
			\centering
			\includegraphics{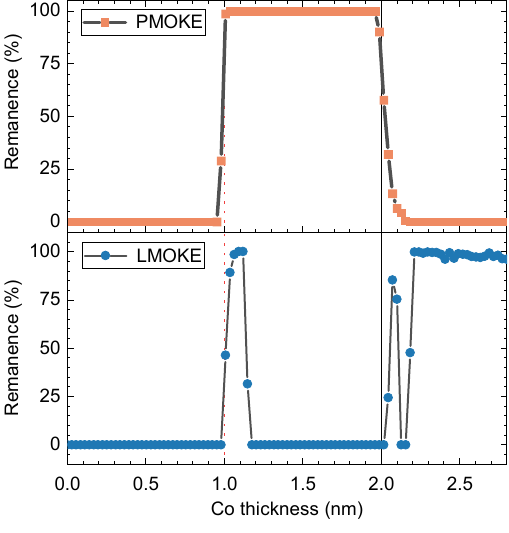}
			\caption{Remanence extracted from the polar MOKE (orange) and longitudinal MOKE (blue) measurement on the same wedge sample as discussed in Fig. \ref{fig:fig2}.}
			\label{fig:fig3}
		\end{figure}

       However, before we can discuss the nature of this change up to large Gd thicknesses, we first need to establish the nature of the $t_\text{OOP}$ and $t_\text{IP}$ transitions at larger Gd thicknesses beyond the PMOKE remanence. To this end, we performed a series of PMOKE and LMOKE measurements on a Ta(4)/Pt(4)/Co(0-3)/Gd(10)/TaN(4)/Pt(2) sample around the two transition thicknesses, as seen for $t_\text{OOP}$ in LMOKE and PMOKE in Figs. \ref{fig:fig2}a and \ref{fig:fig2}b, respectively. At the lowest thickness, we observe no response in either MOKE configuration, suggesting a paramagnetic state (in general, $T_\text{C}=T_\text{room}$ at $t_\text{OOP}$), which transforms to an OOP state with increasing thickness. This contrasts to the behavior seen in LMOKE (Fig. \ref{fig:fig2}c) and PMOKE (Fig. \ref{fig:fig2}d) around $t_\text{IP}$. Here, we observe a transition from an OOP oriented system at low $t_\text{Co}$ through an OOP multidomain state to an IP-oriented system at large $t_\text{Co}$.  To map these different transitions to the actual thickness of the Co layer throughout the wedge, we extract the remanence at every location from full hysteresis loops and plot them in Fig. \ref{fig:fig3}. Here, it can be clearly seen that even at these large Gd thicknesses, the paramagnetic regime at low $t_\text{Co}$ transitions directly into the OOP regime. In both cases, at the transitions, we observe a mixed response in the LMOKE and PMOKE signals, which we attribute to the anisotropy axis not being well defined around $t_\text{OOP}$ and $t_\text{IP}$, combined with a potential small out-of-plane component of the magnetic field.

    \subsection{VSM-SQUID characterization}
\begin{figure}
			\centering
			\includegraphics{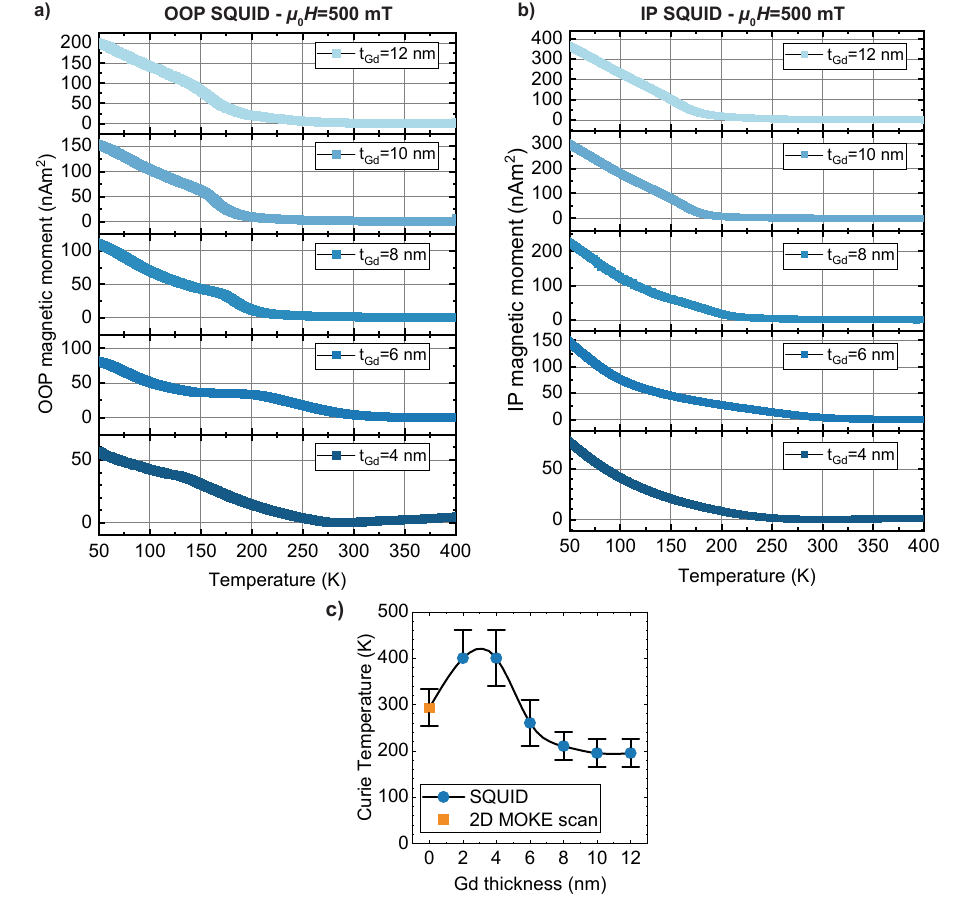}
			\caption{Out-of-plane (a) and in-plane (b) SQUID measurements of a series of Ta(4)/Pt(4)/Co(0.7)/Gd(x)/TaN(4)/Pt(2) samples at an applied magnetic field of 500 mT along the measurement direction. c): Curie temperature extracted from the OOP SQUID measurement as a function of the Gd-thickness, the solid line is a guide to the eye. The orange data point is extracted from the onset of magnetic ordering in the phase diagram in Fig. \ref{fig:fig1}c (i.e., $t_\text{OOP}$ at $t_\text{Gd}=$0 nm).}
			\label{fig:fig6}
		\end{figure}

       Thus, given that both the LMOKE and PMOKE signal vanish below $t_\text{OOP}$, the gradual increase of $t_\text{OOP}$ indicates a decrease in $T_\text{C}$ of the stack as more Gd is added. To further support this claim of a changing $T_\text{C}$ with Gd thickness, we performed a set of IP and OOP VSM-SQUID measurements on a set of full sheet Ta(4)/\-Pt(4)/\-Co(0.8)/\-Gd($t_\text{Gd}$)/\-TaN(4)/\-Pt(2), where $t_\text{Gd}\in \left\{4\:,6,8,10,12 \text{nm}\right\}$. We plot the OOP and IP SQUID response of the magnetic moment as a function of temperature of the different samples in Fig. \ref{fig:fig6}a and \ref{fig:fig6}b, respectively. 
       
       Focusing on the OOP SQUID response in Fig. \ref{fig:fig6}a, we observe two distinct types of behavior: for $t_\text{Gd}=4$ nm, the sample is magnetically ordered across the entire range (just as for $t_\text{Gd}=2$ nm, not shown), with a compensation point at $T=280$ K. This matches in general with the phase diagram (Fig. \ref{fig:fig1}c), where at $t_\text{Gd}=4$ nm the sample is expected to be ferromagnetic and close to the Gd-dominated regime. At $t_\text{Gd}=6$ nm, we observe a marked shift in the response, with the magnetization only starting to set in at $\sim 270$ K. This decrease in $T_\text{C}$ continues when going to $t_\text{Gd}=8$ nm and $t_\text{Gd}=10$. The $T_\text{C}$ values that follow from these measurements are shown in Fig. \ref{fig:fig6}c. Interestingly, the observed trend qualitatively matches the dependence of $t_\text{OOP}$ on the thickness of Gd observed in the MOKE data in Fig. \ref{fig:fig1}c, showing an initial increase of $T_\text{C}$ (or a decrease of $t_\text{OOP}$) followed by a subsequent decrease as the thickness of Gd increases.

       Our observation that the magnetic properties of the Co / Gd bilayers still change significantly with increasing Gd thickness aligns with the findings of Sala \textit{et al.}\cite{Sala2022Hall} in the bilayer system. They did not observe the shift in $T_\text{C}$, likely due to the relatively large thickness of the Co film, but observed a consistent reduction of the net moment with increasing Gd thickness, although it is likely that the interface is no longer affected at the investigated thicknesses. The origin of the reduction of $T_\text{C}$ is not yet fully understood. One hypothesis is that as the Gd becomes thicker, the induced magnetization of the Gd increases, and the phase transition starts to become more dominated by the Gd. Similar phenomena are known to occur in ultrathin heterogeneous thin-film materials once dimensions approach the typical range of the spin-spin correlation length \cite{Bovensiepen1998,Wang1992,Zhang2001}. Generally speaking, a link between the magnitude of the proximity-induced magnetization and $T_\text{C}$ is consistent with earlier results from Koyama \textit{et al.} when studying the Pt/Co/Pt system \cite{Koyama2015}, although they importantly reported an increase in $T_\text{C}$ with increasing induced moment compared to the decrease observed in our work.
       
       The experimental results presented so far, along with those in the literature, provoke us to try to understand the underlying cause of the changes in the magnetic properties of these synthetic ferrimagnetic systems with varying Gd thickness. We believe that two main effects play a role in explaining the observed behavior: sputter-induced intermixing and changes in $T_\text{C}$ due to finite-size effects in the Gd.

    \subsection{Growth kinetics}
       We can investigate the sputter-induced intermixing experimentally by manipulating the sputtering process through the pressure of the Ar working gas. With increasing pressure of the working gas, the number of collisions between the sputtered atoms and the working gas increases, leading to a stronger degree of thermalization of the typical kinetic energy of the sputtered atoms due to a greater amount of collision with the background gas \cite{Meyer1981}. Therefore, we performed PMOKE on a set of Ta(4)/Pt(4)/Co(0.7)/Gd(0-6)/TaN(4)/Pt(2) samples, where the Gd-wedged layer was deposited at different working pressures: 0.05, 0.01, and 0.001 mBar, to investigate the effect on the magnetostatics. With decreasing pressure, we expect the Co/Gd interface roughness to increase as a result of the increased kinetic energy upon impact, leading to more magnetic moment in the Gd due to the increased number of Co-Gd nearest neighbors.

\begin{figure}
    		\centering
    		\includegraphics{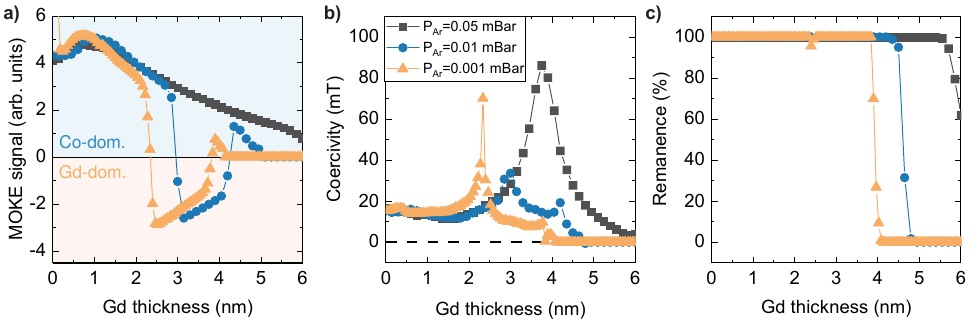}
    		\caption{PMOKE characterization of a Pt/Co/Gd(wedge) where the Gd is deposited at different working gas pressures. Different metadata extracted from the hysteresis loops are plotted: a): MOKE signal at positive saturation. b): Coercive field. c): Remanence.}
    		\label{fig:fig4}
    	\end{figure}

       Hysteresis loops as a function of the Gd thickness were measured across the samples, from which we extracted the MOKE signal at positive saturation (Fig.\ \ref{fig:fig4}a), the coercive field (Fig. \ref{fig:fig4}b), and the remanence (Fig. \ref{fig:fig4}c) as a function of the Gd thickness. For the stack structure discussed here, we can use the onset of compensation as a gauge for the balance between the Co and Gd moments, as shown in Fig. \ref{fig:fig4}a. If we follow the MOKE signal for the 0.01 mBar sample (blue), we first note that at $t_\text{Gd}=0$ nm the MOKE signal is positive, indicating that the magnetic moment is dominated by the Co. At approximately 2.5 nm, the MOKE signal changes sign, indicating that the Gd moment dominates the response to the applied field in the underlying hysteresis loops. Compensation is also indicated by the peak in coercivity at the compensation point, where the MOKE signal switches from positive to negative.

       As the thickness is further increased, we first see a switch back to the Co-dominated regime before the PMOKE signal vanishes completely. As discussed earlier, the disappearance of the MOKE signal occurs when the $T_\text{C}$ of the stack drops below room temperature. The switch from Gd-dominated to Co-dominated just before the transition to a paramagnetic state is not fully understood. We speculate that around $T_\text{C}$, the induced moment in the Gd no longer scales linearly with the moment in the Co, leading to the switch. This may be similar to the vanishing of the proximity-induced magnetization at the ferromagnet/Pt interface near $T_\text{C}$ observed by Inyang \textit{et al.} \cite{Inyang2019}.

       Comparing the MOKE signal (Fig. \ref{fig:fig4}a) for the 0.01 mBar sample to the 0.001 mBar sample (orange), we observe that both the transition to the Gd-dominated regime and the transition to the paramagnetic regime are shifted toward lower Gd thicknesses. We attribute this to the increased intermixing at the Co/Gd interface, as this would increase the average amount of Co-Gd nearest neighbors. In contrast, the 0.05 mBar sample, where we would expect sputter-induced intermixing effects to be diminished, does not display a compensation point in the range studied. However, from the significant increase in coercivity at around 4 nm (Fig. \ref{fig:fig4}b), we can infer that this sample is also very close to compensation in that region.

       In general, the trend between the working gas pressure and the magnetostatic response of our Co/Gd systems is clear and suggests that sputter-induced intermixing can play a significant role in these kinds of system. We note that this trend also reproduces if, instead of the pressure during the deposition of the Gd, we alter the pressure of the capping layer deposition (for more discussion, also on the possible contribution of oxidation during the growth process, see App. \ref{app:Cap}). To support the claim that sputter-induced intermixing can significantly impact the magnetostatic response of the Co/Gd system, and to explore how and why the deposition of nominally paramagnetic Gd at large thicknesses (i.e., many Gd atomic layers separated from the Co/Gd interface) affects the system’s magnetostatics, we will now conduct a numerical study of the deposition process.

    \section{Modeling}
    
        \subsection{Growth kinetics}
\begin{figure}
    		\centering
    		\includegraphics{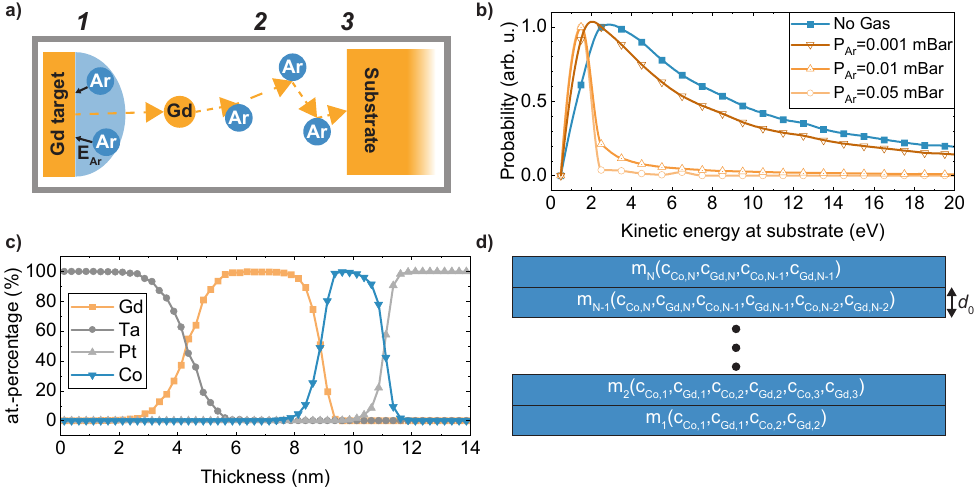}
    		\caption{a): Schematic of the three main steps in the sputter deposition process modeling: Collisions of Ar-ions with the Gd-target, thermalization of sputtered atoms through collisions with the Ar gas, and deposition on the substrate. b): Normalized energy distribution of sputtered atoms for different working gas pressures in the sputtering chamber. c): Typical atomic density profile after simulation of the Co, Gd, and Ta layers deposition. d): Schematic of the atomic density profile being divided into different layers, used as input for the layered Weiss model.}
    		\label{fig:fig5}
    	\end{figure}

        In Fig. \ref{fig:fig5} we show an illustration of the modeling approach that is used to estimate the quantitative change in magnetic moment and effective anisotropy due to sputter-induced intermixing effects are shown in. For our discussion, we first revisit three key characteristics of the thin-film growth process during magnetron sputter deposition, illustrated in Fig. \ref{fig:fig5}a. 1: The sputtered atoms are ejected from the target because of the high kinetic energy of ions in the Ar plasma ignited near the target material. 2: The sputtered atoms travel through the Ar working gas in the sputtering chamber, reducing their average kinetic energy before reaching the substrate. 3: Impact of the sputtered atoms on the substrate, ultimately leading to the formation of a thin film. 

        The first two steps were addressed using the software package Stopping and Range of Ions in Matter (SRIM) \cite{ZIEGLER2010}. These simulations follow the binary collision approximation, which describes sputtering by a sequence of independent binary collisions between atoms. Some important approximations include the following: the Gd target is amorphous, and no dynamic changes in the target are considered. Despite these approximations, the program has been shown to be suitable for determining key factors that impact the interface in layered systems \cite{Eberl2013, Sakhonenkov2022}. This approach was used to determine the energy distributions of Co, Ta, and Gd atoms sputtered from the target by high-energy Ar$^+$ ions. For all materials, we used the default parameters of SRIM for the displacement energy, lattice energy, and surface energy of the atomic species. For example, for Gd, this led to the purple energy distribution in Fig. \ref{fig:fig5}b, which closely resembles the typical Thompson distribution for energies in a sputtering process \cite{Thompson1986}. Ultimately, the energy distribution of the atoms reaching the substrate is mainly determined by the compound probability of an atom being ejected with a particular energy and the remaining energy of an ejected atom when it reaches the substrate.

        To estimate the latter quantity, we simulated the transmission of Gd atoms from the target location to the substrate. The main parameters for these simulations are the sample-to-target distance (10 cm in our system), the density and species of the working gas through which the atoms are transmitted, and the kinetic energy of the Gd atoms. We calculate the gas density $\rho$ using the ideal gas law: $\rho=P/(RT)$, $P$ is the absolute pressure of the gas, $R=208.07$ $\text{J/(kg}\cdot\text{K)}$ is the specific gas constant for Ar, and $T=293$ K is the absolute gas temperature, assumed to be room temperature. For experimental pressures of 0.05, 0.01, and 0.001 mBar, we find densities of $8.2\cdot10^{-8}$, $1.6\cdot10^{-8}$ and $1.6\cdot10^{-9}$ $\text{g/cm}^3$, respectively. We note that especially for a pressure of 0.05 mBar, some degree of gas heating can be expected \cite{Rossnagel1988,Kolev2008}. However, since for this pressure the sputtered atoms are found to be thermalized regardless of the temperature correction, we decided to keep the same temperature in the calculation of the density. We simulate the transmission of 20,000 atoms through the gas column for every Gd, Co, and Ta atom kinetic energy from 1 to 200 eV with a spacing of 1 eV in energy space. The upper limit of 200 eV is derived from the discharge voltage of around 200 V that accelerates the Ar$^+$ ions, with hundreds of eV being a typical value for the incident energy of plasma ions \cite{Gudmundsson2020}. However, we note that, in reality, a distribution of energies is responsible for the sputtering process. SRIM registers the number and kinetic energy of the transmitted atoms, from which we can calculate the probability $P(E_0,E)$ of a sputtered atom with a specific kinetic energy after ejection from the target $E_0$, reaching the end of the gas column with a remaining kinetic energy $E$. The total probability density of an atom with energy $E$, $p(E)$, can then be calculated using the probability of an atom being ejected with energy $E_0$ during the sputtering process, $p_s(E_0)$:
        \begin{equation}
            p(E)=\sum_{E_0=1 \, \text{eV}}^{200 \, \text{eV}} p_s(E_0) p_t(E_0,E).
        \end{equation}

        The results of these calculations for the three pressures are also plotted in Fig. \ref{fig:fig5} b, where we normalize the distributions with respect to the maximum probability density. Generally, an increase in pressure leads to a decrease in the average energy of incident atoms on the samples, as expected. These simulations provide an estimate of how we expect the energy distribution of the sputtered atoms to change due to the ambient gas pressure. This allows us to use kinetic simulations of the deposition process to understand the changes in the sharpness of the Co/Gd interface and the subsequent changes in the net magnetic moment and effective anisotropy in the next section.

        TRIDYN simulations are used to simulate interface formation using the probability distribution obtained from the SRIM simulation \cite{Moller1984}. The TRIDYN code is also based on the binary collision approximation but allows for a dynamic simulation of irradiation and deposition processes \cite{Moller1988}. We first simulate the deposition of the Co layer on top of Pt over a range of thicknesses, and then we simulate the deposition of Gd with varying thicknesses using the profiles obtained at different sputtering pressures. In this manner, it is possible to numerically reproduce the double-wedge samples. To simulate the intermixing effects introduced by the capping layer, we model the deposition of 4 nm of Ta on top of Gd.

        We carry out these TRIDYN simulations for Co thicknesses between 0 and 2 nm and Gd thicknesses between 0 and 5 nm, with spacings of 0.25 nm. The final step in the analysis is to calculate the magnetostatics of the stack given the calculated atomic density profiles, as shown in Fig. \ref{fig:fig5}c. We do this using the layered Weiss model, which is described in detail elsewhere \cite{Beens2019Intermixing}, though a brief derivation as well as the chosen parameters can be found in Appendix \ref{app:Parameters}. 

        \begin{figure}
    		\centering
    		\includegraphics{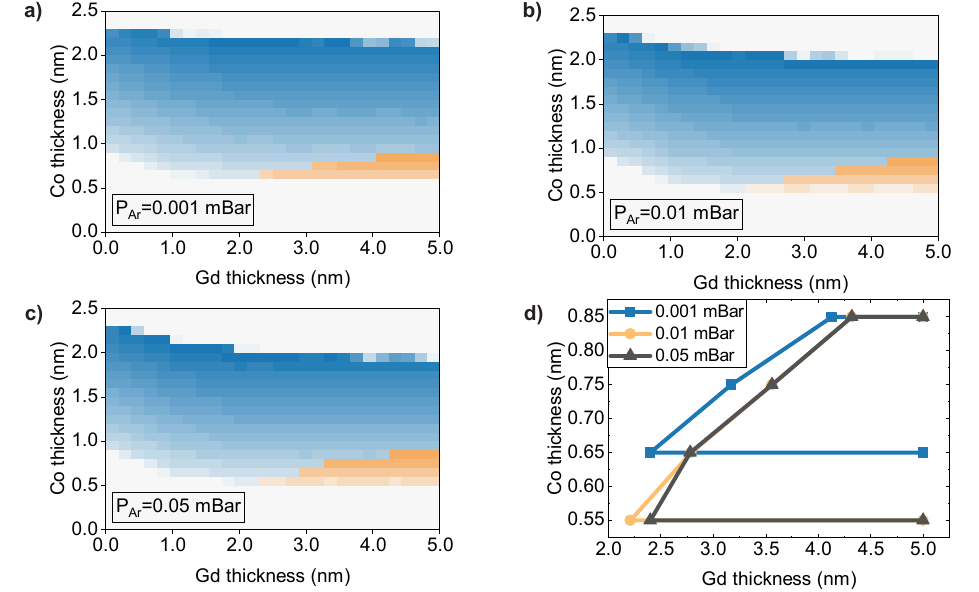}
    		\caption{Numerically calculated phase diagram based on the SRIM and TRIDYN simulations of the deposition process, and the layered Weiss model for Ar working gas pressures of 0.001 mBar (a), 0.01 mBar (b), and 0.05 mBar (c). Blue: OOP Co-dominant. Orange: OOP Gd-dominant. White: No OOP remanence (i.e., in-plane magnetization for large $t_\text{Co}$, $M=0$ around $t_\text{Co}=0$ nm). d): Outline of the Gd-dominated regions in the phase diagram for the different pressures.}
    		\label{fig:fig7}
    	\end{figure}
     
        The end result of these calculations is shown in Fig. \ref{fig:fig7}, where we present the numerically calculated phase diagram for Ar pressures of 0.001, 0.01, and 0.05 mBar in Figs. \ref{fig:fig7}a, \ref{fig:fig7}b, and \ref{fig:fig7}c, respectively. We also plotted the outline of the Gd-dominated region of the different phase diagrams in Fig. \ref{fig:fig7}d. There are several important observations we can make based on these phase diagrams, in comparison with the experimental phase diagram in Fig. \ref{fig:fig1}c. First, our model reproduces the decrease of $t_\mathrm{OOP}$, the transition thickness from paramagnetic to OOP, at low Gd thicknesses. This is caused by the boost in exchange energy that accompanies the change of the Co/Ta interface to a Co/Gd interface, which is also reflected in the actual atomic density profiles (see Appendix \ref{app:tridyn}). We also note that, similar to the experiments with different growth pressures presented in Fig. \ref{fig:fig4}, we observe magnetization compensation and a shift to increased Gd magnetization upon decreasing the Ar pressure. This further supports the hypothesis that sputter-induced intermixing is a relevant factor in the magnetostatics of this type of system, which follows from the very strong sensitivity of the induced moment to the quality of the interface and the extremely thin Co layer. 

        We also clearly observe the limitations of such an approach based solely on sputter-induced intermixing and the layered Weiss model. The most striking difference with the experiment is that the increase in $t_\text{OOP}$ and $t_\text{IP}$ at high Gd thicknesses is not reproduced in this model description. This follows from the fact that the Weiss model only takes into account constant nearest-neighbor exchange and therefore does not reflect the finite-size effects (beyond those caused by the reduced number of nearest neighbors at the interface) and the effect of spin waves on the critical phenomena governing the phase transition and, by extension, the amount of induced magnetization in the Gd, which earlier work has found to be quite important in the ultrathin film regime \cite{Baberschke2008,Wang1992}. This is a topic we address in the next section when we discuss our analytical model for proximity-induced magnetization. Another very important point is the choice of the definition of the anisotropy energy, which is not unambiguous. We have shown in previous work that the typical models used for the description of the effective anisotropy are not adequate to describe this type of magnetic system \cite{Kools2022}. In particular for the transition from OOP to IP, this can affect the shape of the calculated phase diagram at $t_\text{IP}$ and cause a better reflection of the experimental results. Nevertheless, this modeling approach still provides important insight into the magnitude and range of sputter-induced intermixing effects on the magnetic moment for this type of ultrathin synthetic ferrimagnet. Given that it is likely a relevant effect based on our simulations and experiments, it may prove to be another interesting way to tune the properties of these systems.

  \subsection{Analytical model}
   Considering that the layered Weiss model is not suitable for describing our experiments in the high-thickness Gd regime, we will further explore our understanding of the system through a minimal analytical model of the proximity-induced magnetization at the Co/Gd interface, taking into account finite-size effects and changes in $T_\text{C}$. Importantly, this is not intended to be a complete model of the system, as many extrinsic and intrinsic effects determine the ultimate magnetic state of the system. Therefore, the main focus of this section will be to demonstrate that the main trends in the phase diagram presented in Fig. \ref{fig:fig1} can be reasonably explained based on such a description.
    
For this purpose, we closely follow a description previously used by Lim \textit{ et al.} and Omelchenko \textit{et al.} to describe proximity-induced magnetization at a Py / Pt interface \cite{Omelchenko2018,Lim2013}. Analogously, with more details in Appendix \ref{app:ParametersAn}, we derive the following expression for the Gd magnetization, $M_\text{Gd}$, as a function of the distance to the Co/Gd interface, $x$,

\begin{equation}\label{eq: Interm2}
    M_\text{Gd}(x)=M_0 e^{-x/\lambda},
\end{equation}

where $\lambda=\sqrt{\chi A}$ is the characteristic decay length of proximity-induced magnetization, defined by the Gd susceptibility, $\chi$ and exchange stiffness $A$, and $x=0$ corresponds to the position of the Co/Gd interface where the thickness of the Gd develops along positive $x$. The induced magnetization at the Co/Gd interface $M_0$ is assumed to originate from the exchange coupling, $M_0=J_\text{ex}M_\text{Co}\chi$, where $\mu_0J_\text{ex}M_\text{Co}$ is the effective exchange field on the Gd at the Co/Gd interface. We also note that with increasing thickness, $T_\text{C}$ of Gd is expected to increase in the thickness range we consider due to finite-size effects \cite{Farle1993}, and assume that $A$ follows the same scaling law $f_\text{ex}(t_\text{Gd})$ proposed by Zhang and Willis: $f_\text{ex}(t_\text{Gd})=1/(1+c_2/t_\text{Gd})^\gamma$ \cite{Zhang2001}, where $c_\text{2}$ and $\gamma$ are phenomenological parameters that describe these finite-size effects. 

Combining these with eq. \eqref{eq: Interm2}, and evaluating the integral across the whole Gd thickness, the total Gd moment per unit area $m_\text{Gd}$ can be calculated to be:
\begin{eqnarray}
   m_{Gd}= &&\int_{0}^{t_\text{Gd}}  M_0 e^{-x/\lambda} \,\mathrm{d}x = \nonumber\\
   &&J_\text{ex}\chi M_\text{Co} \lambda_0 \sqrt{f_\text{ex}} \left[1-e^{-t_\text{Gd}/\left(\lambda_0 \sqrt{f_\text{ex}}\right)}\right].
\end{eqnarray}

  We describe the spontaneous moment per unit area of the Co by means of the Curie-Weiss law
  \begin{equation}\label{eq: Comom}
      m_\text{Co}(T)=M_\text{Co}d_{Co}=M_\text{s,0}d_\text{Co}\left(1-\frac{T}{T_\text{C}}\right)^\beta,
  \end{equation}

where $M_\text{s,0}$ is the saturation magnetization when approaching 0 K, $d_\text{Co}$ is the Co thickness, $T_\text{C}$ is the Curie temperature, and $\beta$ is the critical exponent. We can calculate the net magnetic moment $m_\text{net}$ as the difference between $m_\text{Co}$ and $m_\text{Gd}$

\begin{eqnarray}
    m_\text{net}=&&M_{s,0}\left(1-\frac{T}{T_\text{C}}\right)^\beta \nonumber \\
    &&\left[d_{Co} -  J_\text{ex}\chi\lambda_0 \sqrt{f_\text{ex}} \left[1-e^{-t_\text{Gd}/\left(\lambda_0 \sqrt{f_\text{ex}}\right)}\right]\right].
\end{eqnarray}

In order to then describe the three regimes in the magnetostatic phase diagrams, we also need to describe the effective anisotropy and Curie temperature of our synthetic ferrimagnet as a function of the Co and Gd thickness. For the effective anisotropy, we consider two competing contributions. The first is a perpendicular anisotropy contribution from the Co/Pt interface $K_\mathrm{OOP}$, which is described as
\begin{equation}
    K_\text{OOP}=\frac{K_\mathrm{s}}{(t_\text{Co}+t_\text{Gd,eff})},
\end{equation}
where $K_\text{s}$ is the interfacial anisotropy strength, and $t_\text{Gd,eff}=t_\text{Gd}$ for $t_\text{Gd}<2\lambda_0$ and $t_\text{Gd,eff}=2\lambda_0$ for $t_\text{Gd}\geq2\lambda_0$, reflects the fact that the thickness relevant for the effective anisotropy calculated is partially determined by the Gd, although where to put this boundary is not well-defined given the inhomogeneity of the interface and the decaying magnetization in the Gd. For the shape anisotropy $K_\text{IP}$, we use the following expression
\begin{equation}
    K_\text{IP}=\frac{1}{2}\mu_0\left(\frac{m_\text{net}}{(t_\text{Co}+t_\text{Gd,eff})}\right)^2,
\end{equation}
where the term within parentheses represents the magnetization of the film. The effective anisotropy of the layer is then given as $K_\text{eff}=K_\text{OOP}-K_\text{IP}$. If this quantity is positive (negative) the magnetization is taken to be OOP (IP).

Next, we will address $T_\text{C}$ of our material. As we have observed in the experimental part of this work, $T_\text{C}$ of the Co/Gd bilayer depends in a non-trivial way on the thickness of the Gd- and Co-layer. We model $T_\text{C}$ first considering that $T_\text{C}$ of our synthetic ferrimagnet scales linearly with the Co thickness within the ultrathin regime that we consider, following the theoretical description by Zhang and Willis \cite{Zhang2001} and the experimental results from Schneider \textit{et al.} \cite{Schneider1990}. To have a continuous description of the Gd thickness-dependence of $T_\text{C}$, we fit a regular Gaussian to the experimental data in Fig. \ref{fig:fig6}c, which is given as

\begin{equation}\label{eq:gauss}
    T_\text{C}(t_\text{Gd})=T_0+\sqrt{\frac{4\ln\left(2\right)A^2}{\pi w^2}}\exp\left(\frac{-4\ln\left(2\right)\left(t_\text{Gd}-t_\text{Gd,c}\right)^2}{w^2}\right).
\end{equation}

 Finally, we model $T_\text{C}$ of the full stack by combining the Co- and Gd-thickness dependence we extracted from the experiments into the final expression

\begin{equation}\label{Eq:Tc}
    T_\text{C}=a_0 t_\text{Co} - \left(T_\text{C}(t_\text{Gd})-T_\text{C}(0))\right),
\end{equation}

where $a_\text{0}=293\,\text{K}/0.7 \,\text{nm}=418$ K/nm is a parameter describing the dependence of $T_\text{C}$ on thickness. As a benchmark for this value, we used the experimental observation that the transition from paramagnetic to ferromagnetic in the phase diagram at room temperature (293 K) occurs at $t_\text{Co}=0.7$ nm. The latter term in eq. \eqref{Eq:Tc} indicates the difference between the experimentally observed $T_\text{C}$ with and without Gd. We note that the description given by Eq. \eqref{Eq:Tc} breaks down in the limit of very small Co-thicknesses (<0.1 nm), possibly giving a negative Curie temperature. Although for such small amounts of Co, the Curie temperature will inevitably be below room temperature, we will just consider these regions as paramagnetic.

\begin{figure}
    		\centering
    		\includegraphics{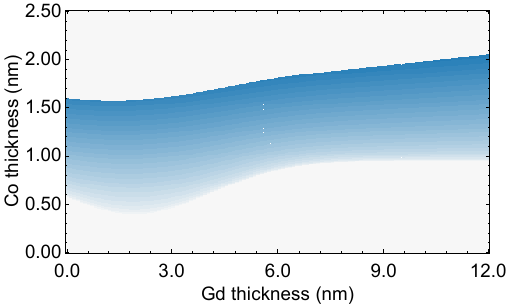}
    		\caption{Phase diagram extracted from our analytical model of the proximity induced magnetization with changing exchange strength due to the changing Gd-thickness. Blue indicates Co-dominated OOP magnetization, whereas white indicates no OOP remanent mangetization. }
    		\label{fig:fig8}
\end{figure}

We plot a phase diagram obtained through this model in Fig. \ref{fig:fig8}, using the parameters listed in Table \ref{param: Analytical}. For the chosen parameters, which we will discuss in more detail below, we observe that the earlier phase diagram (Fig.\ref{fig:fig1}d) reproduces quite well. The change in the induced magnetization that accompanies the increase in the effective exchange within the Gd as the thickness of the layer increases and the associated reduction in effective anisotropy as the net magnetization is lowered explains the experimental phase diagram in the large Gd-thickness regime quite well. For the calculation of this phase diagram, the values of $y_0$, $t_\text{Gd,c}$, $A$, and $w$ are best-fit parameters of the fit of eq. \eqref{eq:gauss} to the data in Fig. \ref{fig:fig6}c. The value for $K_\text{s}$ is a typical value for the interfacial anisotropy at the Co/Pt interface for our sputtered systems \cite{Kools2022}, and we choose $\beta$ = 0.4 as a typical value for the critical exponent for Co \cite{Vaz2008}. For the parameters $c_\text{2}$ and $\gamma$, which determine the finite-size effects on the effective exchange parameter of Gd, we follow the characterization by Zhang and Willis \cite{Zhang2001}, who find $\gamma=2.8$, but we adjust $c_\text{2}=9$ nm such that the scaling law matches the fraction of the bulk $T_\text{C}$ observed in our experiments at $t_\text{Gd}=12$ nm. We attribute the difference between our experiments and those reported by Zhang \textit{et al.} \cite{Zhang2001}, and later Farle \textit{et al.} \cite{Farle1993}, to the difference between our disordered sputtered samples to their monocrystalline Gd layers. We take $\lambda_0=0.3$ nm after the results of Lim and Omelchenko \cite{Lim2013, Omelchenko2018}, though noting that the actual typical decay length in Co/Gd has not been established, and is generally hard to define due to interfacial intermixing and (subsequent) inhomogeneity. This also holds for $J_\text{ex}\chi$, which determines the magnitude of the interfacially induced magnetization, and the resulting phase diagram is characterized by a strong correlation between the choice of $\lambda_0$ and $J_\text{ex}\chi$. We find a good correspondence between the experiment and the model for $J_\text{ex}\chi=-1$, and this value is reasonably consistent with our earlier work \cite{Kools2022}.

Although we note that the choices of $\lambda_0$ and $J_\text{ex}\chi$ are within reasonable bounds compared to earlier experiments regarding proximity-induced magnetization, their exact values are currently unknown and can significantly affect the model results. However, the fact that the shift in $T_\text{C}$ in the stack and the expected increased effective exchange due to finite-size effects in the Gd can quantitatively explain the main trends in the phase diagram using parameters from similar experiments suggests that these kinds of effects indeed play an important role in understanding this type of ultrathin synthetic ferrimagnet. Ultimately, relatively simple models like the two presented in this and the previous section will likely never be able to fully explain the complexity of these ultrathin systems. Integrating the effect of changes in every interface due to finite-size effects, changes in the crystallinity of the materials, and intermixing in a single model will, on the other hand, lead to an overparameterized model that could fit any set of data. Still, these reduced models serve as a useful test environment to further our understanding of the typical scaling and magnitudes of the different effects.

\begin{table}
\caption{Model parameters used in the generating of Fig. \ref{fig:fig8}.}
\begin{tabular}{|l@{\qquad}|l@{\qquad}|l@{\qquad}|}
\hline
\textbf{Symbol} & \textbf{Value} & \textbf{Unit}    \\ \hline
$\gamma$             &  2.8          &  -          \\ \hline
$c_\text{2}$          &  20           &  nm         \\ \hline
$J_\text{ex}\chi$     &  -1           &  -          \\ \hline
$\lambda_0$           &  0.6          &  nm         \\ \hline
$\beta$               &  0.4          &  -          \\ \hline
$K_\text{s}$          &  1.0          &  mJ/m$^2$   \\ \hline 
$a_0$                 &  418          &  K/nm       \\ \hline 
$T_0$                 &  195          &  K          \\ \hline
$t_\text{Gd,c}$       &  2.59         &  nm         \\ \hline
$A$                   &  1225         &  nm$\cdot$K \\ \hline
$w$                   &  5.3          &  nm         \\ \hline
\end{tabular}
\label{param: Analytical}
\end{table}

\section{Conclusion}
In summary, we have demonstrated the strong sensitivity of $T_\text{C}$ of the Co/Gd bilayer to the thickness of the Gd layer, even at thicknesses where the deposited Gd is separated from the Co/Gd interface by several atomic layers, and up to about 10 nm. The presence of Gd can severely influence the Curie temperature of the stack, even when the interface between Co and Gd is no longer expected to be impacted, owing to finite-size effects and the subsequent changing exchange interaction strength in the Gd. To explain the rich behavior observed in the PMOKE remanence and SQUID characterization, we presented two separate models related to sputter-induced intermixing and the increasing interface-induced magnetization in the Gd due to finite-size effects. These models could explain different aspects of the experiments. Further research may focus on extending the methods to other ultrathin 3d-4f synthetic ferrimagnets and investigating the (damped) oscillatory dependence of the effective anisotropy with Gd thickness. We conclude by noting that the insights garnered in this work provide important guidelines towards the understanding and development of materials that could become essential for integrated magneto-photonics, through the relation between $T_\text{C}$ and the threshold fluence for all-optical switching, and the emerging field of orbitronics.

\section*{Acknowledgements}
    This work was part of the research program Foundation for Fundamental Research on Matter (FOM) and Gravitation program “Research Center for Integrated Nanophotonics,” which are financed by the Dutch Research Council (NWO). This work was supported by the Eindhoven Hendrik Casimir Institute (EHCI). This project has received funding from the European Union’s Horizon 2020 research and innovation program under the Marie Skłodowska-Curie grant agreement No 861300.
		
\newpage
\appendix

\section{Additional 2D-wedge measurement}\label{app:Longrangewedge}
\begin{figure} 
			\centering
			\includegraphics{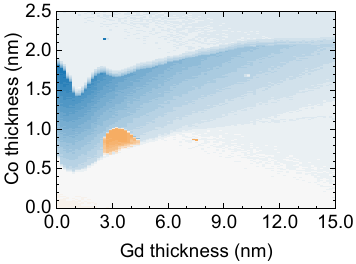}
			\caption{Polar MOKE remanent signal scan on a  Ta(4 nm)/Pt/\-Co(0-2.5)/\-Gd(0-15)/\-TaN(4)/\-Pt(2) sample after saturation with a 1 T OOP field.}
			\label{fig:figapp2}
		\end{figure}

In Fig. \ref{fig:figapp2}, we plot the PMOKE remanence of a 2D-wedge sample of the following composition: Ta(4 nm)/Pt/\-Co(0-2.5)/\-Gd(0-15)/\-TaN(4)/\-Pt(2). This sample was measured analogously to the one presented in the main text in Fig. \ref{fig:fig1}c. The range in which the change saturates coincides with the typical range at which finite size effects on the Curie temperature of Gd thin films saturate \cite{Farle1993,Oshea1999}.
\newpage
  \section{Capping layer pressure and oxidation}\label{app:Cap}
\begin{figure}
    		\centering
    		\includegraphics{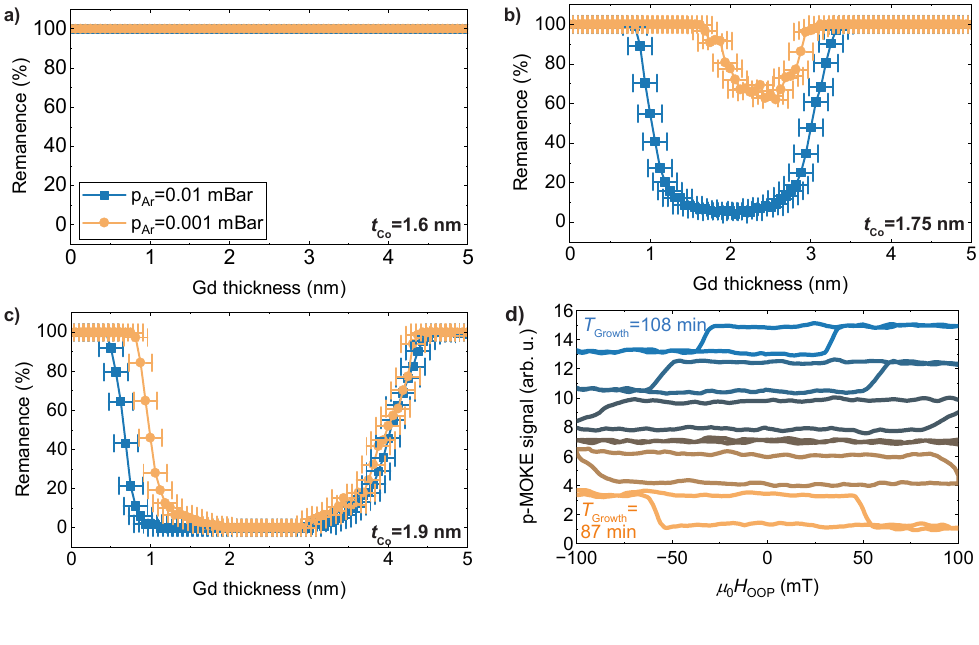}
    		\caption{a,b,c): PMOKE remanence extracted from hysteresis loops measured on a Ta(4)/Pt(4)/Co(1.6(a)/1.75(b)/1.9(c))/Gd(x)/Ta(4) for two different growth pressures during the deposition of the Ta caping layer. d): In-situ PMOKE characterization of a Ta(4 nm)/\-Pt(4)/\-Co(0.65)/\-Gd(1.7)/\-Co(0.7)/\-Gd(1.5).}
    		\label{fig:App1}
    	\end{figure}

In Fig. \ref{fig:App1} we show the PMOKE remanence extracted from hysteresis loops of two Ta(4)/\-Pt(4)/\-Co(1.6(a)/1.75(b)/1.9(c))/\-Gd(x)/\-Ta(4) samples where the only difference is the deposition pressure of the Ta capping layer deposited to protect the Gd from oxidation. Compared to the sample discussed in the article, we do not use TaN, as changing the pressure of the Ar will inevitably also alter the composition of the TaN layer, complicating the analysis. Therefore, we have chosen atomic Ta as the capping layer. When we consider the remanence, we observe that the high (low) pressure sample transitions from an OOP to IP orientation at lower (higher) Co thickness, suggesting a smaller (larger) effective anisotropy. This is consistent with the explanation of sputter-induced intermixing, since additional intermixing at the Co/Gd interface induced by the Ta would lead to a decrease in the net magnetic moment due to the increased Gd magnetization and hence to a decrease in the shape anisotropy leading to an OOP orientation for larger Co thicknesses. 

We note that although oxidation due to impurities in the working gas may have a similar effect on the magnetic balance, the purity of the working gas combined with the very short exposure to the working gas and the extensive presputtering make this an unlikely contribution \cite{Wandelt1985,Cohen2011,Cohen2013}, although it cannot be completely excluded. To further substantiate this claim, we performed in-situ PMOKE measurements on a Ta(4)/\-Pt(4)/\-Co(0.65)/\-Gd(1.7)/\-Co(0.7)/\-Gd(1.5) sample, without a capping layer, which has a compensation point at a top Gd-layer thickness of about 0.8 nm. We measured PMOKE hysteresis loops in a vacuum chamber at a pressure of 8$\cdot$10$^{-9}$ mBar at different times $T_\text{growth}$ after the deposition, which are plotted in Fig. \ref{fig:App1}d. We observe the transition from Gd- to Co-dominated, indicated by the change in sign of the hysteresis loop. As can be seen, this transition takes place only at $T_\text{growth}\approx 90$ minutes, and although it is by no means certain that oxidation takes place linearly with time, it further suggests that oxidation effects due to the limited oxygen exposure from contaminations in the working gas are likely going to be small.

\newpage

\section{Details layered Weiss model simulations}\label{app:Parameters}

The magnetization, normalized to $M_\text{S}$, $m_{i,\text{Co(Gd)}}$ in each sublayer $i$ depends on the local exchange splitting $\Delta_{i,\text{Co(Gd)}}$ and spin quantum number $S_\text{Co(Gd)}$
        \begin{equation}
	       m_{i,\text{Co(Gd)}} = \frac{1}{S_\text{Co(Gd)}} \sum\limits_{m_s=-S_\text{Co(Gd)}}^{S_\text{Co(Gd)}}m_s f_{m_s}
        \end{equation}
        with, 
        \begin{equation}
	       f_{m_s} = \frac{\mathrm{e}^{-\Delta_{i,\text{Co(Gd)}} m_s/(k_\mathrm{B}T)}}{\sum\limits_{m_s}\mathrm{e}^{-\Delta_{i,\text{Co(Gd)}} m_s/(k_\mathrm{B}T)}}.
        \end{equation}
        Here, $k_\text{B}$ is the Boltzmann constant, $T$ is the absolute temperature. A more detailed expansion of the exchange splitting for a (111)-textured facet yields:
       \begin{align}
           \Delta_\text{i,Co}=\frac{1}{4}\Delta^\text{bulk}_\text{i-1,Co}+\frac{1}{2}\Delta^\text{bulk}_\text{i,Co}+\frac{1}{4}\Delta^\text{bulk}_\text{i+1,Co}\\           
           \Delta_\text{i,Gd}=\frac{1}{4}\Delta^\text{bulk}_\text{i-1,Gd}+\frac{1}{2}\Delta^\text{bulk}_\text{i,Gd}+\frac{1}{4}\Delta^\text{bulk}_\text{i+1,Gd},    
       \end{align}
           where we introduced the bulk exchange splitting for a Gd$_{1-x_\text{i}}$Co$_{x_i}$ alloy with a Co concentration $x_i$ \cite{Beens2019Intermixing,Beens2019Alloy}:

        \begin{align}
               \Delta^\text{bulk}_\text{Co,i}=x_i\gamma_\text{Co-Co}m_\text{i,Co}+\left(1-x_i\right)\gamma_\text{Co-Gd}m_\text{i,Gd}\\
               \Delta^\text{bulk}_\text{Gd,i}=x_i\gamma_\text{Gd-Co}m_\text{i,Co}+\left(1-x_i\right)\gamma_\text{Gd-Gd}m_\text{i,Gd}.
        \end{align}
        Here, $\gamma_{kl}=j_{kl}zD_\text{s,l}S_l$ ($k,l\in{\text{Co,Gd}}$) is a measure for the exchange coupling strength, where $D_\text{s}=\mu_\text{at}/2S$ is the number of spins per lattice site, $j_{kl}$ is the exchange coupling constant, and $z$ the numbers of nearest neighbors.

        We extract the atomic concentrations $x_i$ from the TRIDYN simulations described before, which are binned into "layers" of $d_\mathrm{0}=0.25~\mathrm{nm}$ and then calculate the total net magnetization in each layer $M_{i,\text{net}}$ through:
        \begin{equation}
            M_{i,\text{net,A}}= x_i M_\mathrm{s,Co}m_{i,\text{Co}} + (1-x_i) M_\mathrm{s,Gd}m_{i,\text{Gd}}.
        \end{equation}
        Another important quantity we can calculate using the net magnetic moment is the surface normalized effective OOP anisotropy, $K_\text{eff,A}$. This follows from the competition between the shape anisotropy and the contribution from the Pt/Co interface \cite{Kools2022}:
        \begin{equation}
            K_\text{eff,A}= K_\text{s} - \frac{1}{2}\mu_\text{0}\sum_i d_\mathrm{0} M_{i,\text{net}}^2,
        \end{equation}
        where $\mu_\text{0}$ is the magnetic permeability of vacuum and $K_\text{s}$ is the interfacial OOP anisotropy strength. The values of all the parameters used for the numerical modeling of the growth process and subsequent intermixing can be found in table \ref{param:SRIM}.
\begin{table}
\caption{Overview of parameter values used in the SRIM-TRIDYN-Layered Weiss modeling effort.}
\begin{tabular}{|l@{\qquad}|l@{\qquad}|l@{\qquad}|}
\hline
\textbf{Symbol} & \textbf{Value} & \textbf{Unit} \\ \hline
 $S_\text{Co}$             &  1/2           &  $\hbar$     \\ \hline
 $S_\text{Gd}$             &  7/2           &  $\hbar$      \\ \hline
 $\mu_\text{at,Co}$        &  1.72          &  $\mu_\mathrm{B}$  \\ \hline
 $\mu_\text{at,Gd}$        &  7.55          &  $\mu_\mathrm{B}$  \\ \hline
 $K_\text{s}$              &  1.0  &  $\mathrm{mJ}\cdot\mathrm{m}^{-2}$ \\ \hline
 $z$                       &  12             &  -          \\ \hline
 $j_\text{Co-Co}$          &  86.1           &  meV          \\ \hline
 $j_\text{Co-Gd}$          &  -33.4          &  meV          \\ \hline
 $j_\text{Gd-Co}$          &  -21.2          &  meV         \\ \hline
 $j_\text{Gd-Gd}$          &  15.2           &  meV          \\ \hline
 $T$                       &  295            &  K              \\ \hline
 $k_\text{B}$              &  $1.38\cdot10^{-23}$        & $\mathrm{J}\cdot\mathrm{K}^{-1}$              \\ \hline
 $\mu_\text{0}$            &       $1.26\cdot10^{-6}$ & $\mathrm{N}\cdot\mathrm{A}^{-2}$ \\ \hline
 $\mu_\text{B}$            &       $9.27\cdot10^{-24}$ & $\mathrm{J}\cdot\mathrm{T}^{-1}$ \\ \hline
 $\hbar$            &       $1.05\cdot10^{-34}$ & $\mathrm{J}\cdot\mathrm{s}$ \\ \hline
\end{tabular}
\label{param:SRIM}
\end{table}

\newpage
\section{Calculated intermixing profiles as a function of pressure}\label{app:tridyn}
\begin{figure}
			\centering
			\includegraphics{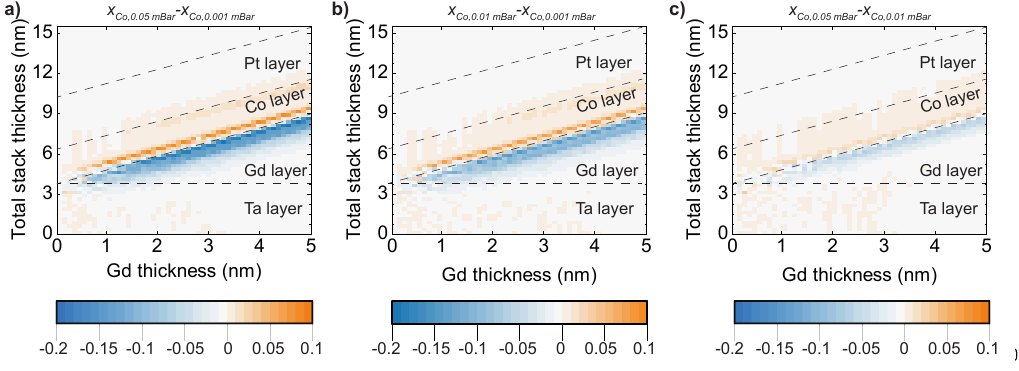}
			\caption{Difference in Co atomic fraction, $x$, as a function of Gd thickness, obtained from the Tridyn simulations as a function of Gd thickness for different Ar working gas pressures used in the simulations of the deposition process. Different layers based on their nominal layer thicknesses are indicated. a): $x_{Co,0.05 \,\text{mBar}}-x_{Co,0.001 \,\text{mBar}}$. b): $x_{Co,0.01 \,\text{mBar}}-x_{Co,0.001 \,\text{mBar}}$. c): $x_{Co,0.05 \,\text{mBar}}-x_{Co,0.01 \,\text{mBar}}$}
			\label{fig:figapp3}
		\end{figure}

\begin{figure}
			\centering
			\includegraphics{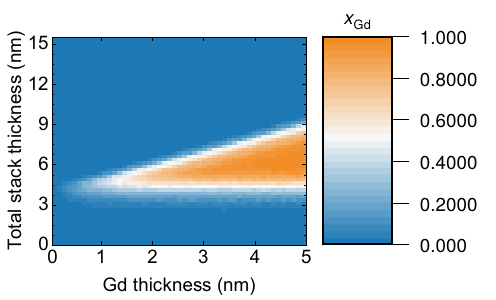}
			\caption{Gd atomic concentration distribution as a function of the deposited Gd-thickness as calculated with the Tridyn simulations.}
			\label{fig:figapp4}
		\end{figure}

The key output from the Tridyn and SRIM simulations of the growth process of a Pt/Co/Gd/Ta stack described in the main text are the atomic fraction of the Co and Gd for a given Ar working gas pressure during the Gd deposition, $P$, $x_{\text{Co,}P}$ and $x_{\text{Gd,}P}$ respectively. A useful metric to image the change in the atomic profiles at the Co/Gd interface with the pressure of the working gas during the deposition of Gd is the difference in Co concentration $\Delta x_\text{Co,P1,P2}=x_\text{Co,P1}-x_\text{Co,P2}$. In Fig. \ref{fig:figapp3}a, \ref{fig:figapp3}b, and \ref{fig:figapp3}c we plot $\Delta x_\text{Co,0.05\, mBar,0.001 \, mBar}$, $\Delta x_\text{Co,0.01\, mBar,0.001 \, mBar}$, and $\Delta x_\text{Co,0.05\, mBar,0.01 \, mBar}$, respectively, at a Co thickness of 2.5 nm. The main difference in the atomic fraction occurs at the Co/Gd interface, as this is the main interface that would be affected by a change in $P$. The blue and orange bands bordering the nominal position of the Co/Gd interface in the simulations indicate that with decreasing pressure, the interface becomes less sharp. The differences in the Co atomic fraction range from about 15-20\% in Fig. \ref{fig:figapp3}a to around 5\% in Fig. \ref{fig:figapp3}c. We also plot the $x_\text{Gd,0.001 \, mBar}$ in Fig. \ref{fig:figapp4}, from which we can learn that the typical roughness of the Ta / Gd interface is around 1 to 1.5 nm within the simulation, which lies at the root of the gradual decrease in $t_\text{OOP}$ at low Gd thicknesses in Fig. \ref{fig:fig7}, and suggests that a similar effect may be responsible for the experimental observation.

\newpage

\section{Details analytical model}\label{app:ParametersAn}
At the basis of this model lies the following phenomenological free energy density function, which originates from Ginzburg-Landau theory, in the absence of an applied magnetic field \cite{landau2013,Murata1972}
\begin{equation}
    f(x)=\frac{\mu_0}{2\chi}M_{Gd}(x)^2 + \frac{\mu_0 A}{2}\left[\diff{M_{Gd}(x)}{x}\right]^2,
\end{equation}
which we employ here to describe the induced magnetization in paramagnetic Gd, $M_\text{Gd}$. Here, $\mu_0$ is the magnetic permeability of vacuum, $\chi$ is the magnetic susceptibility, $A$ is the exchange stiffness and $m$ is the local induced moment. The reader is reminded that $T_\text{C}$ of pure Gd in the thickness range of interest in this work is well below room temperature \cite{Farle1993}.  We assume that the local moment only varies meaningfully in the direction along the thin-film normal, which we define to be the x-direction with $x=0$ at the Co/Gd interface. Note that we do not take into account intermixing within this model. The area density of the total free energy related to the proximity-induced moment in the Gd is then given by
\begin{equation}
    F= \int_{0}^{d_\text{Gd}} \left(\frac{\mu_0}{2\chi}M_\text{Gd}(x)^2 + \frac{\mu_0 A}{2}\left[\diff{M_\text{Gd}(x)}{x}\right]^2\right) \,\mathrm{d}x,
\end{equation}

where $d_\text{Gd}$ is the thickness of the Gd layer. We can use the variational principle to show that the magnetization is described by the following differential equation

\begin{equation}\label{eq:diffeq}
    \frac{\mu_0}{\chi} M_\text{Gd}(x)-\mu_0 A \diffp[2]{M_\text{Gd}(x)}{x}.
\end{equation}

Subsequently, we can define the boundary conditions for solving eq.\ \eqref{eq:diffeq}; we assume $M_\text{Gd}(0)=M_0$ at the Co/Gd interface and $M_\text{Gd}(x)=0$ as $x$ tends to infinity. The solution to Eq. \eqref{eq:diffeq} given these boundary conditions is

\begin{equation}\label{eq: Interm1}
    M_\text{Gd}(x)=M_0 \exp\left(-\frac{x}{\lambda}\right),
\end{equation}
which is the starting point of the discussion in the main text.

\clearpage

	\bibliography{Bibliography.bib}

	\end{document}